\documentclass[twocolumn,journal]{IEEEtran}

\usepackage{amsfonts}
\usepackage{amsmath}
\usepackage{amsthm}
\usepackage{amssymb}
\usepackage{graphicx}
\usepackage[T1]{fontenc}
\usepackage{supertabular}
\usepackage{longtable}
\usepackage[usenames,dvipsnames]{color}
\usepackage{bbm}
\usepackage{fancyhdr}
\usepackage{breqn}

\usepackage{capt-of}
\usepackage{tikz}
\usetikzlibrary{matrix}
\usepackage{endnotes}
\usepackage{soul}
\usepackage{marginnote}
\newcommand{\mathsym}[1]{}
\newcommand{\unicode}[1]{}

\usepackage{colortbl}

\usepackage[framemethod=TikZ]{mdframed}
\usepackage[framemethod=TikZ]{mdframed}
\usepackage{framed}

\mdfsetup{%
	outerlinewidth=1,skipabove=20pt,backgroundcolor=yellow!30, outerlinecolor=black,innertopmargin=0pt,splittopskip=\topskip,skipbelow=\baselineskip, skipabove=\baselineskip,ntheorem,roundcorner=5pt}

\mdtheorem[nobreak=true,outerlinewidth=1,
backgroundcolor=yellow!50, outerlinecolor=black,innertopmargin=0pt,splittopskip=\topskip,skipbelow=\baselineskip, skipabove=\baselineskip,ntheorem,roundcorner=5pt,font=\itshape]{result}{Result}

\mdtheorem[nobreak=true,outerlinewidth=1,
backgroundcolor=yellow!30, outerlinecolor=black,innertopmargin=0pt,splittopskip=\topskip,skipbelow=\baselineskip, skipabove=\baselineskip,ntheorem,roundcorner=5pt,font=\itshape]{theorem}{Theorem}

\mdtheorem[nobreak=true,outerlinewidth=1,
backgroundcolor=gray!10, outerlinecolor=black,innertopmargin=0pt,splittopskip=\topskip,skipbelow=\baselineskip, skipabove=\baselineskip,ntheorem,roundcorner=5pt,font=\itshape]{remark}{Remark}

\mdtheorem[nobreak=true,outerlinewidth=1,
backgroundcolor=gray!10, outerlinecolor=gray!10,innertopmargin=0pt,splittopskip=\topskip,skipbelow=\baselineskip, skipabove=\baselineskip,ntheorem,roundcorner=5pt,font=\itshape]{definition}{Definition}

\mdtheorem[nobreak=true,outerlinewidth=1,
backgroundcolor=pink!30, outerlinecolor=black,innertopmargin=0pt,splittopskip=\topskip,skipbelow=\baselineskip, skipabove=\baselineskip,ntheorem,roundcorner=5pt,font=\itshape]{quaestio}{Quaestio}

\mdtheorem[nobreak=true,outerlinewidth=1,
backgroundcolor=yellow!50, outerlinecolor=black,innertopmargin=5pt,splittopskip=\topskip,skipbelow=\baselineskip, skipabove=\baselineskip,ntheorem,roundcorner=5pt,font=\itshape]{background}{Background}

\mdtheorem[nobreak=true,outerlinewidth=1,
backgroundcolor=gray!10, outerlinecolor=black,innertopmargin=5pt,splittopskip=\topskip,skipbelow=\baselineskip, skipabove=\baselineskip,ntheorem,roundcorner=5pt,font=\itshape]{nothing}{}

\mdtheorem[nobreak=true,outerlinewidth=1,
backgroundcolor=pink!50, outerlinecolor=black,innertopmargin=5pt,splittopskip=\topskip,skipbelow=\baselineskip, skipabove=\baselineskip,ntheorem,roundcorner=5pt,font=\itshape]{point}{Point}
\mdtheorem[nobreak=true,outerlinewidth=1,
backgroundcolor=pink!50, outerlinecolor=black,innertopmargin=5pt,splittopskip=\topskip,skipbelow=\baselineskip, skipabove=\baselineskip,ntheorem,roundcorner=5pt,font=\itshape]{lemma}{Lemma}

\mdtheorem[nobreak=true,outerlinewidth=1,
backgroundcolor=pink!50, outerlinecolor=black,innertopmargin=5pt,splittopskip=\topskip,skipbelow=\baselineskip, skipabove=\baselineskip,ntheorem,roundcorner=5pt,font=\itshape]{commentary}{Comment}

\mdtheorem[nobreak=true,outerlinewidth=1,
backgroundcolor=pink!50, outerlinecolor=black,innertopmargin=5pt,splittopskip=\topskip,skipbelow=\baselineskip, skipabove=\baselineskip,ntheorem,nobreak=true,roundcorner=5pt,font=\itshape]{proposition}{Proposition}

\begin{document}

\title{{\color{Red}
What You See and What You Don't See:\\
 The Hidden Moments of a Probability Distribution}}

\author{Nassim Nicholas Taleb\IEEEauthorrefmark{1}\IEEEauthorrefmark{2}\\     
   \IEEEauthorblockA{  \IEEEauthorrefmark{1} American University of Beirut \IEEEauthorrefmark{2} 
   Universa Investments}
   \thanks{  November 25, 2026.
     Thanks to Zhuo Xi, participants at "Heavy Tails 2019" in Eindhoven in November 2019, Bert Zwart, Paul Embrechts, Wim Schoutens, and others (Nidal Selmi, Armand D'Angour, Nassim Dehouche). This revision corrects typos in the earlier version spotted by AI (Claude, ChatGPT).
   }
   }

\maketitle
\begin{mdframed}
\smallskip
   \section{Abstract/Introduction}
Empirical distributions	have their in-sample maxima as natural censoring. We look at the "hidden tail", that is, the part of the distribution in excess of the maximum for a sample size of $n$. Using extreme value theory, we examine the properties of the hidden tail and calculate its moments of order $p$.

The method is useful in showing how large a share of the true statistical mean (or higher moments) lies, for a given $n$, beyond what is visible in sample; it is considerable for $\alpha$ close to 1.

We believe that the method has applications to numerous domains, particularly finance and insurance where relatively small samples cause chronic underestimations and overestimations of the risks and expected returns.

\end{mdframed}

\begin{figure} 
 	\includegraphics[width=\linewidth]{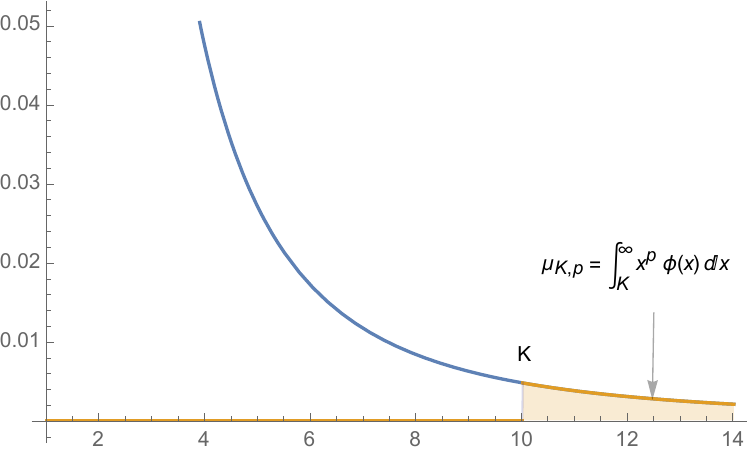}
\caption{The $p^{th}$ moment above $K$, the maximum in-sample observation.}\label{meanaboveK}
 \end{figure}
\section{The Invisible Tail for a Power Law}
Consider $K_n$ the maximum of a sample of $n$ independent identically distributed variables in the power law class; $K_n=\max \left(X_1,X_2,\ldots,X_n\right)$. Let $\phi(.)$ be the density of the underlying distribution. We can decompose the moments in two parts, with the "hidden" moment above $K_n$, as shown in Fig. \ref{meanaboveK}:

$$\mathbbm{E}(X^p)=\underbrace{\int_L^{K_n} x^p \phi (x) \, dx}_{\mu_{L,p}}
+\underbrace{\int_{K_n}^{\infty } x^p \phi (x) \, dx}_{\mu_{K,p}}$$
where $\mu_{L,p}$ is the observed part of the distribution and $\mu_{K,p}$ the hidden one. We note that $\phi(.)$ is not rescaled.


\begin{figure} 
 	\includegraphics[width=\linewidth]{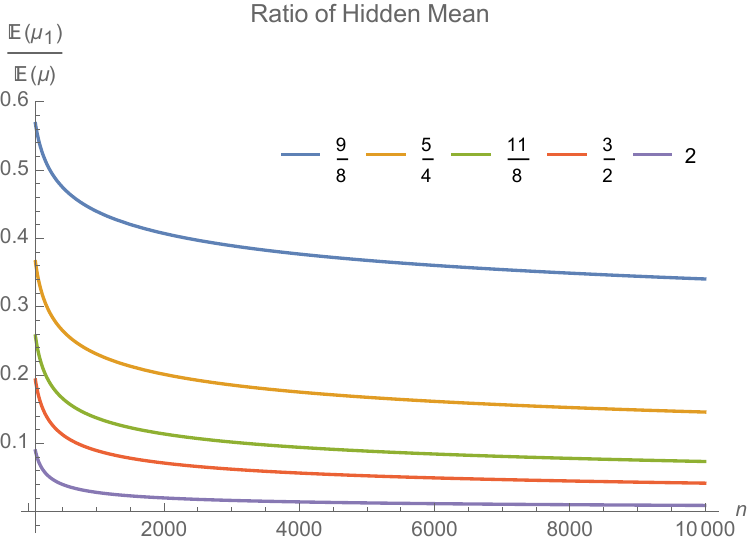}
\caption{Proportion of the hidden mean in relation to the total mean, as a function of $n$, for different parametrizations of the tail exponent $\alpha$.}\label{ratio1}
 \end{figure}

\begin{figure} 
 	\includegraphics[width=\linewidth]{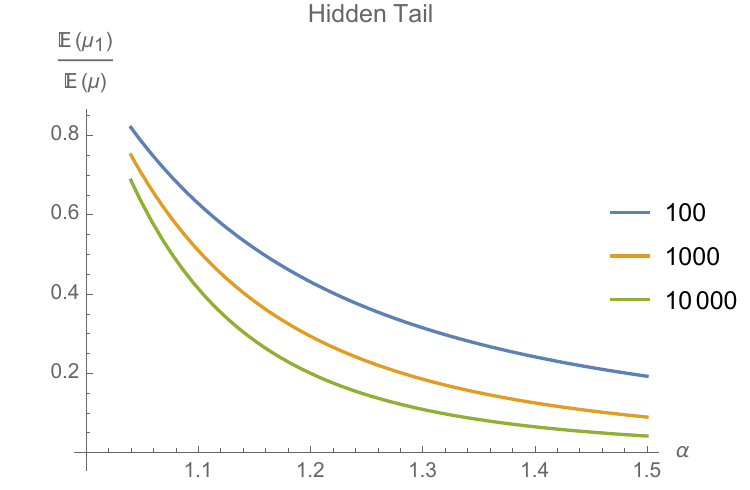}
\caption{Proportion of the hidden mean in relation to the total mean, as a function of $\alpha$, for different sample sizes $n$.}\label{ratio2}
 \end{figure}

\begin{proposition}
Let $K^*$ be a point above which the survival function of the random variable $X$ can be satisfactorily approximated by a constant times a power, that is, $\mathbbm{P}(X>x) \approx L^{\alpha} x^{-\alpha}$ for $x \geq K^*$.

Under the assumption that $K_n>K^*$, the hidden $p^{th}$ moment $\mu_{K,p}$ for $n$ observations has for (asymptotic) density $g_{n,p,\alpha}(.)$:
\begin{dmath}
	g_{n,p,\alpha}(z)=n L^{\frac{\alpha  p}{p-\alpha }} \left(z-\frac{p z}{\alpha }\right)^{\frac{p}{\alpha -p}} \exp \left(-n L^{\frac{\alpha  p}{p-\alpha }}
	 \left(z-\frac{p z}{\alpha }\right)^{\frac{\alpha }{\alpha-p }}\right) \label{density}
\end{dmath}
for  $z\geq 0$, $\alpha>p$, and $L>0 $.
\end{proposition}
In other words $\mu_{K,p}$ is approximately Weibull distributed for a large sample, with shape parameter $\frac{\alpha}{\alpha-p}$. The mean becomes
\begin{equation}
\mathbbm{E}(\mu_{K,p}) \approx L^p n^{\frac{p}{\alpha }-1} \Gamma \left(1-\frac{p}{\alpha }\right), \label{hiddenmean}
\end{equation}
and, since $\mathbbm{E}(X^p)\approx \frac{\alpha L^p}{\alpha-p}$ for the Pareto, the hidden share of the $p^{th}$ moment is free of the scale $L$:
\begin{equation}
\frac{\mathbbm{E}(\mu_{K,p})}{\mathbbm{E}(X^p)}=\Gamma \left(2-\frac{p}{\alpha }\right) n^{\frac{p}{\alpha }-1}, \label{hiddenratio}
\end{equation}
which, for $p=1$, is the quantity shown in Figs. \ref{ratio1} and \ref{ratio2}.

The proof is as follows. The $p^{th}$ moment above $K$, with $K\geq L>0$, can be derived as

\begin{equation}
	\mu_{K,p}=	\frac{\alpha  L^{\alpha } K^{p-\alpha }}{\alpha -p}\;, \;\alpha>p \label{muK}
\end{equation}
and we need to calculate its distribution.

For the full distribution $g_{n,p,\alpha}(z)$, take a Pareto with scale $L$, so $K_{min}=L$. The maximum has CDF $\left(1-L^{\alpha}K^{-\alpha}\right)^n \approx e^{-n L^{\alpha}K^{-\alpha}}$, so $K_n \sim \mathcal{F}(\alpha, L n^{\frac{1}{\alpha}})$, a Fr\'echet distribution with PDF: $f_K(K)= \alpha  n K^{-\alpha -1} L^{\alpha } e^{-n \left(\frac{L}{K}\right)^{\alpha }}$. Since \eqref{muK} is decreasing in $K$,
\begin{align*}
\mathbbm{P}\left(\mu_{K,p}\leq z\right)&=\mathbbm{P}\left(K_n\geq L^{\frac{\alpha}{\alpha-p}}\left(z-\tfrac{pz}{\alpha}\right)^{-\frac{1}{\alpha-p}}\right)\\
&=1-e^{-nL^{\frac{\alpha p}{p-\alpha}}\left(z-\frac{pz}{\alpha}\right)^{\frac{\alpha}{\alpha-p}}},
\end{align*}
and differentiating with respect to $z$ gives the required result. The mean \eqref{hiddenmean} follows from $\mathbbm{E}(K_n^{p-\alpha})=L^{p-\alpha}n^{\frac{p}{\alpha}-1}\Gamma\left(2-\frac{p}{\alpha}\right)$.

For an exact Pareto no asymptotics are needed: $L^{\alpha}K_n^{-\alpha}$ is the minimum of $n$ uniforms on $(0,1)$, hence Beta$(1,n)$ distributed, and
\begin{equation}
\mathbbm{E}(\mu_{K,p})=L^p\, \Gamma\left(1-\frac{p}{\alpha}\right)\frac{\Gamma(n+1)}{\Gamma\left(n+2-\frac{p}{\alpha}\right)}, \label{hiddenmeanexact}
\end{equation}
of which \eqref{hiddenmean} is the large $n$ limit. (The exact support is $0<z\leq\frac{\alpha L^p}{\alpha-p}$; the Fr\'echet approximation leaves a mass $e^{-n}$ on $K_n<L$.)

More generally, for any continuous distribution, by exchangeability a new observation $X_{n+1}$ exceeds $K_n$ with probability $\frac{1}{n+1}$, in which case it is the maximum $K_{n+1}$ of $n+1$ observations; hence
\begin{equation}
\mathbbm{E}(\mu_{K,p})=\frac{\mathbbm{E}\left(K_{n+1}^p\right)}{n+1}, \label{identity}
\end{equation}
of which \eqref{hiddenmeanexact} is the Pareto case. The identity also holds in the extended sense for nonnegative random variables with infinite moments.

\begin{remark}
For $p=0$ our result recovers a classical fact: the survival function evaluated at the sample maximum has, for large $n$, an exponential distribution with pdf
\begin{equation}
g_{n,0,\alpha}(z)=n e^{-n z},
\end{equation}
which depends only on $n$. Exceedance probability does not depend on the thickness of the tails. Exactly, by the probability integral transform, it is Beta$(1,n)$ distributed, with pdf $n(1-z)^{n-1}$ on $[0,1]$ and mean $\frac{1}{n+1}$, for \emph{any} continuous distribution.
\end{remark}

\begin{remark}
When $\alpha>1$ the sample mean remains unbiased in expectation: the deficit is made up by the rare samples in which the maximum is very large. The hidden mean measures the part of the true mean that the sample at hand cannot reveal; its counterpart is that the sample mean lies below the true mean with probability exceeding $\frac{1}{2}$, markedly so for $\alpha$ close to $1$ (about $0.8$ for $\alpha=\frac{5}{4}$, $n=1000$, by simulation).
\end{remark}

\begin{figure} 
 	\includegraphics[width=\linewidth]{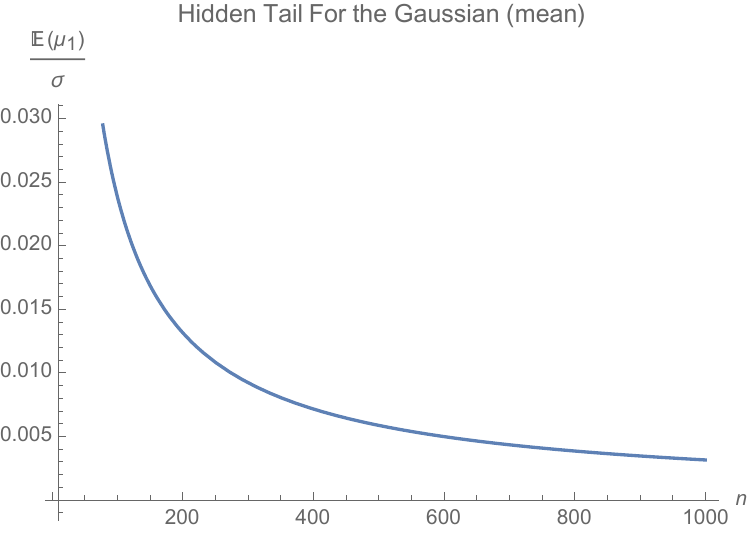}
\caption{The hidden mean in relation to the standard deviation for the Gaussian, as a function of $n$.}\label{gaussianfig}
 \end{figure}

\section{Comparison with other distributions}

To get the expectation where closed forms are not available (say with the Lognormal), we just need to get the integral with a stochastic lower bound $K>K_{min}$  
$$\int_{K_{min}}^\infty\left(\underbrace{\int_{K}^{\infty } x^p \phi (x) \, dx}_{\mu_{K,p}}\right)f_K(K)dK.$$


For a Gaussian with PDF $\phi^{(g)}(.)$ indexed by $(g)$, and $K\geq 0$,
$$\mu_{K,p}^{(g)}=\int_K^\infty x^p \phi^{(g)}(x) dx=\frac{2^{\frac{p}{2}-1} \Gamma \left(\frac{p+1}{2},\frac{K^2}{2}\right)}{\sqrt{\pi }}.$$
Rather than go through the Gumbel, to which the Gaussian maximum converges very slowly, it is preferable to use the exact distribution of the maximum, from the CDF of the Standard Gaussian $F^{(g)}$, with $K_{min}=-\infty$:
$$f_K^{(g)}(K)=\frac{\partial F^{(g)}(K)^n}{\partial K}=\frac{e^{-\frac{K^2}{2}} 2^{\frac{1}{2}-n} n \,\text{erfc}\left(-\frac{K}{\sqrt{2}}\right)^{n-1}}{\sqrt{\pi }},$$
where erfc is the complementary error function.

For $p=0$, the expectation of the "invisible tail" is $\frac{1}{n+1} \approx \frac{1}{n}$, since, with $u=F^{(g)}(K)$:
\begin{dmath}
	\int_{-\infty}^{\infty } \frac{e^{-\frac{K^2}{2}} 2^{-n-\frac{1}{2}} n \,\text{erfc}\left(\frac{K}{\sqrt{2}}\right) \left(\text{erf}\left(\frac{K}{\sqrt{2}}\right)+1\right)^{n-1}}{\sqrt{\pi} }\, dK =\int_0^1 n (1-u) u^{n-1} \, du =\frac{1}{n+1}
\end{dmath}
in accordance with Remark 1.

For $p=1$, $\mu_{K,1}^{(g)}=\phi^{(g)}(K)$ for all $K$, and the hidden mean is $\int_{-\infty}^{\infty}\phi^{(g)}(K) f_K^{(g)}(K)\, dK$, shown in Fig. \ref{gaussianfig}. It is not apparently possible to obtain it (or higher moments) in closed form, but it behaves as $\frac{\sqrt{2 \log n}}{n}$ for large $n$ (numerically $0.0249$ for $n=100$ and $0.0032$ for $n=10^3$): a rapid decline, faster than the $n^{-\frac{1}{2}}$ rate of convergence of the mean of the Gaussian under the law of large numbers \cite{TaStat}. For the Gaussian the hidden tail is immaterial next to ordinary sampling error; for the power law with $\alpha$ near $1$, by \eqref{hiddenratio}, it is not.


\section{Applications and Conclusion}
The property of the hidden moments is useful to understand the "properties of the unseen", in environments where past data provides insufficient evidence --and we are aware of it; hence we can quantify such unknown (or at least get an idea of its magnitude). It thus has an epistemological value  as we can tell beforehand the magnitude of the underestimation, and how confident we can be from past data.

 This is particularly useful for one-tailed fat-tailed distributions where the sample mean will, with high probability, be underestimating the true mean if the tail is to the right, and overestimating it if the tail is to the left. It applies, for instance, to the misestimation of the true P/L of a strategy with long or short volatility profile, to the pricing of options with remote strike prices, to the true expected long term damage from hurricanes and other natural calamities, to the expected level of flooding, to the true properties of war and mean casualties from violence, and many more. Indeed using methods from extreme value theory \cite{deHaan,Embrechts} critically changes the thinking and the conclusions as compared to "evidence based" methods that have statistical flaws under thicker tailed domains.

\flushbottom 


\begin{thebibliography}{99}   
\bibitem{TaStat} N.N. Taleb (2020). \textit{Statistical Consequences of Fat Tails}. STEM Academic Press. 
\bibitem{deHaan} L. de Haan, A. Ferreira (2006). \textit{Extreme Value Theory: An Introduction}. Springer.
\bibitem{Embrechts} P. Embrechts, C. Kl\"uppelberg, T. Mikosch (1997). \textit{Modelling Extremal Events}. Springer.
\end{thebibliography}
\end{document}